\documentclass{article}

\usepackage{latexsym}
\usepackage{amssymb}
\usepackage{amsmath}
\usepackage{cite}

\def\R{{\Bbb R}}
\def\C{{\Bbb C}}

\def\Tr{{\rm Tr}}

\def\T{{\rm T}}

\def\Mat{{\rm Mat}}

\def\diag{{\rm diag}}
\def\cl{{\mathcal C}\!\ell_{1,3}}
\def\Even{{\rm Even}}
\def\Odd{{\rm Odd}}

\def\tr{{\rm tr}}

\newcommand{\be}{\begin{equation}}
\newcommand{\ee}{\end{equation}}
\def\st{\stackrel}
\def\SU{\mathrm{SU}}
\def\U{\mathrm{U}}
\def\su{\mathrm{su}}
\def\u{\mathrm{u}}
\def\O{\mathrm{O}}

\newtheorem{theorem}{Theorem}

\begin{document}

\title{Sketch of a Gauge Model of Gravity \\ with SU(2) Symmetry in Minkowski space}


\author{Nikolay Marchuk\footnote{
Steklov Mathematical Institute, 
Gubkin str. 8, 
Moscow 119991,
Russia\\
email: nmarchuk@mi-ras.ru}}

\maketitle

\begin{abstract}
We propose a gauge model with the SU(2) symmetry, which describes a gravitational interaction of fundamental fermions (leptons and quarks) in the Minkowski space.
In the Standard Model one uses a Dirac-Yang-Mills system of equations with U(2) gauge symmetry for electroweak interactions and with SU(3) gauge symmetry for QCD interactions. A key idea of the model is to use the Dirac-type equation (invented in 2002) instead of the standard Dirac equation. This Dirac-type equation has an additional SU(2) gauge symmetry. The Yang-Mills field, which corresponds to this SU(2) symmetry, we identify with the gravitational field of  interacted fundamental fermions. Some elements of Clifford analysis are used in the model.
\end{abstract}

\medskip

keywords: Clifford algebra, Dirac equation, Gauge symmetry, Gravity, Minkowski space, SU(2) symmetry, Yang-Mills equations











\section*{Introduction}
\label{intro}
It is supposed that a quantum gravity theory must consider a gravitational interaction on the elementary particles level.  In the literature there are many approaches to quantum gravity (see, for example,  \cite{Oriti} and \cite{Rovelli}). In particular,  a class of superstring theories is the favorite in number of publications.

\sloppypar{
A gauge fields approach to quantum gravity was suggested by  R.~Utiyama \cite{Utiyama}
and developed by many researchers ( \cite{Blag, Hay, Hehl, Gia},
 etc.).
}

 In our model, in contrast to the Utiyama model,  we describe the gravitational interaction of fundamental fermions (leptons and quarks) using the unitary gauge group $\SU(2)$. A key idea of the model is to use the Dirac-type equation (see \cite{paper2002}, or \cite{MarchukEng2012} Section 5.8)  with the additional $\SU(2)$ gauge symmetry instead of the standard Dirac equation.

We describe needed mathematical technique in sections 1-3.
In section 4 we propose a system of differential equations for leptons (doublet leptons) with the $\SU(2)\times \U(2)$ gauge symmetry, where the $\U(2)$ symmetry corresponds to the electroweak interaction and the $\SU(2)$ symmetry corresponds to the gravitational interaction of leptons. In section 5 we propose a system of differential equations for quarks with the $\SU(2)\times \U(2)\times \U(3)$ gauge symmetry, where the $\U(2)$ symmetry corresponds to the electroweak interaction, the $\U(3)$ symmetry corresponds to the strong (QCD) interaction, and the $\SU(2)$ symmetry corresponds to the gravitational interaction of quarks (triplet quarks).

In this paper we suggest that the gravitational  field is ``weak'' that means we can neglect a curvature of a space-time manifold and consider particles interactions in the Minkowski space.

In order to describe a ``strong'' gravitational field the considered model must be generalized on a pseudo-Riemannian manifold. In this case we need  one more equation to establish a connection between the gravitational (gauge) field and the metric tensor of a manifold. This will be discussed in a further paper.

There are many questions to be considered for the proposed model -- Lagrangian and Hamiltonian approaches, a quantization, etc.

\section{The Real Clifford Algebra $\cl$ and the Complexified Clifford Algebra $\C\otimes\cl$}
\label{sec1}

Let $\eta=\|\eta^{ab}\|=\|\eta_{ab}\|=\diag(1,-1,-1,-1)$ be the Minkowski matrix and $\cl$ be the Clifford algebra \cite{MarchukEng2012} with generators $e^0,e^1,e^2,e^3$ and with the identity element $e$.   For elements $U,V\in\cl$ an associative product $U,V\to U V$ can be defined using equalities
$$
e^a e^b+ e^b e^a = 2\eta^{ab}e,\quad a,b=0,1,2,3.
$$
The Clifford algebra $\cl$ can be considered as the 16-dimensional real vector space with the following basis numbered by ordered multi-indices of length from $0$ to $4$
\begin{equation}
e,e^0,e^1,e^2,e^3,e^{01},e^{02},e^{03},e^{12},e^{13},e^{23},e^{012},e^{013},e^{023},e^{123},e^{0123},
\label{cl:basis}
\end{equation}
where
$$
e^{a_1\ldots a_k}=e^{a_1}\cdots e^{a_k},\quad\mbox{for}\quad 0\leq a_1<\cdots<a_k\leq4,\ k=2,3,4.
$$
For an integer $k$ ($0\leq k\leq4$) let $\cl^k$ be the vector subspace spanned by the basis vectors from (\ref{cl:basis}) with multi-indices of length $k$. Elements from $\cl^k$ are called {\em grade $k$ elements}. The dimensions of vector spaces $\cl^k$ for $k=0,1,2,3,4$ are equal to $1,4,6,4,1$ respectively.

Now we can define the subspace $\cl^\Even$  of {\em even} elements of the Clifford algebra $\cl$ and the subspace $\cl^\Odd$  of {\em odd} elements of the Clifford algebra $\cl$
$$
\cl^\Even=\cl^0\oplus\cl^2\oplus\cl^4,\quad\cl^\Odd=\cl^1\oplus\cl^3.
$$
Also, in what follows, we use projector operators
\begin{eqnarray*}
\pi^k &:& \cl\to\cl^k,\quad k=0,1,2,3,4;\\
\pi^\Even &:& \cl\to\cl^\Even,\quad \pi^\Odd : \cl\to\cl^\Odd.
\end{eqnarray*}
An arbitrary element $U\in\cl$ can be written in the form
\begin{equation}
U=u e+u_a e^a +\sum_{a_1<a_2}u_{a_1 a_2}e^{a_1 a_2}+\sum_{a_1<a_2<a_3}u_{a_1 a_2 a_3}e^{a_1 a_2 a_3}+u_{0123}e^{0123},\label{decomp:u}
\end{equation}
 where $u,u_a,\ldots,u_{0123}$ are real numbers and we use the convention that $u_a e^a=\sum_{a=0}^4 u_a e^a$.
We have
$$
\pi^0(U)=u e,\quad\pi^1(U)=u_a e^a,\quad\ldots,\quad \pi^4(U)=u_{0123}e^{0123}.
$$
Sometimes for grade $k$ elements we use a notation
\begin{equation}
\st{k}{U} = \pi^k(U),\quad U=\sum_{k=0}^4\st{k}{U}.\label{stkU}
\end{equation}
In what follows we need the linear {\em reverse operator} ${}^\sim : \cl^k\to\cl^k$ such that
$$
(e^{a_1}\cdots e^{a_k})^\sim=e^{a_k}\cdots e^{a_1}.
$$
For an element $U\in\cl$ of the form (\ref{stkU}) we have
$$
U^\sim=\st{0}{U}+\st{1}{U}-\st{2}{U}-\st{3}{U}+\st{4}{U}.
$$
If $U,V\in\cl$, then $(U V)^\sim=V^\sim U^\sim$.

Let us use special notations (without indices) for the two elements of basis (\ref{cl:basis})
$$
\beta = e^0,\quad \theta =e^{0123}.
$$
Note that
$$
\beta^2=e,\quad \theta^2=-e.
$$
Now we define an operator of Hermitian  conjugation for Clifford algebra elements by the formula
$$
U^\dagger := \beta U^\sim\beta.
$$
We see that $(U V)^\dagger = V^\dagger U^\dagger$, $\forall U,V\in\cl$ and
$$
\beta^\dagger=\beta,\quad \theta^\dagger=-\theta.
$$
Also, let us define {\em the trace} of Clifford algbera element $\Tr : \cl\to\R$
$$
\Tr(U) :=\pi^0(U)|_{e\to1}.
$$
Finally, we get {\em a scalar product} of Clifford algebra elements
$$
(U,V):=\Tr(U^\dagger V),\quad U,V\in\cl
$$
such that  $(U,U)>0$,  $\forall U\neq0$.
 Considering the Clifford algebra $\cl$ together with this scalar product, we arrive at
 the 16-dimensional Euclidean space.
\medskip

\noindent{\bf The complexified Clifford algebra $\C\otimes\cl$}. Considering the set of elements of the form (\ref{decomp:u}), where coefficients $u,u_a,\ldots,u_{0123}$ are complex numbers (from $\C$), we arrive at the complexified Clifford algebra $\C\otimes\cl$ with subsets (subspaces) $\C\otimes\cl^\Even$, $\C\otimes\cl^\Odd$, $\C\otimes\cl^k$ ($k=0,1,2,3,4$).

 For elements of $\C\otimes\cl$ we define the operation of complex conjugation
$$
U\to\bar{U}=\bar u e+\bar u_a e^a +\sum_{a_1<a_2}\bar u_{a_1 a_2}e^{a_1 a_2}+\sum_{a_1<a_2<a_3}\bar u_{a_1 a_2 a_3}e^{a_1 a_2 a_3}+\bar u_{0123}e^{0123},
$$
where $\bar u,\bar u_a,\ldots,\bar u_{0123}$ are complex conjugated numbers.

We  consider the (real) Clifford algebra $\cl$ as a subalgebra of the complexified Clifford algebra $\C\otimes\cl$.

Evidently, the projection operators $\pi^k,\pi^\Even,\pi^\Odd$, the trace operator $\Tr$, and the reverse operator ${}^\sim$ can be applied to elements of the complexified Clifford algebra $\C\otimes\cl$.

The operator of Hermitian conjugation of elements of $\C\otimes\cl$ is defined by the formula
$$
U^\dagger =\beta\bar U^\sim\beta.
$$
 Considering the Clifford algebra $\C\otimes\cl$
together with the Hermitian scalar product $(U,V)=\Tr(U^\dagger V)$, we arrive at the 16-dimensional unitary space.

In what follows we use {\em the Hermitian idempotent}
\footnote{Also we may use $\chi =\frac{1}{2}(e + i\theta)$, or $\chi=e$.}
$$
\chi=\frac{1}{2}(e - i\theta)
$$
such that
\begin{eqnarray*}
&&\chi^2=\chi,\quad \chi^\dagger=\chi,\quad \theta \chi=i \chi,\\
 &&[U,\chi]=0,\quad \forall U\in \C\otimes\cl^\Even,
\end{eqnarray*}
where  $[U,V]=U V-V U$.

Let $I(\chi)$ be the left ideal of $\C\otimes\cl$ generated by the idempotent $\chi$, and let
$K(\chi)$ be the intersection of the left and the right ideals
$$
I(\chi)=\{U\in\C\otimes\cl\,:\,U=U \chi\},\quad K(\chi)=\{U\in I(\chi)\,:\,U=\chi U\}.
$$


\section{Some Lie Groups and Lie Algebras}
\label{sec2}
Denote two Lie algebras (w.r.t. the commutator) of Clifford algebra elements by
\begin{eqnarray*}
L_4 &=& \{s\in\cl^\Even\,:\,s^\dagger=-s\},\\
L_3 &=& \{s\in\cl^\Even\,:\,s^\dagger=-s,\ [\beta,s]=0\}.
\end{eqnarray*}
Considering even elements of the basis (\ref{cl:basis}), we see that the Lie algebra $L_4$ is the four-dimensional vector space  spanned by the basis elements $e^{12}$, $e^{13}$, $e^{23}$, $e^{0123}$, and the Lie algebra $L_3$ is the three-dimensional vector space  spanned by the basis elements $e^{12},e^{13},e^{23}$. The Lie algebra  $L_3$ is isomorphic to the Lie algebra $\su(2)$ of anti-Hermitian traceless matrices of second order, and the Lie algebra  $L_4$ is isomorphic to the Lie algebra $\u(2)$ of anti-Hermitian  matrices of second order (note that  $\u(2)$ is isomorphic to    $\u(1)\oplus\su(2)$,  where $\u(1)$ is the one-dimensional Lie algebra spanned by the basis element $\theta=e^{0123}$).

If we denote
$$
\tau^1 = e^{23}, \quad \tau^2=-e^{13},\quad \tau^3=e^{12},
$$
then each  element of $L_3$ can be written as a sum (over $k=1,2,3$)
$$
s=s_k\tau^k,\quad\mbox{where}\quad s_1,s_2,s_3\in\R.
$$

Consider the Lie group (w.r.t. the Clifford product)
$$
G_3 = \{S\in\cl^\Even\,:\, S^\dagger S=e,\ [\beta,S]=0\}.
$$
This Lie group is isomorphic to the Lie group $\SU(2)$ of special unitary matrices of second order. The Lie algebra $L_3$ is the real Lie algebra of the Lie group $G_3$. Any element $S$ of the Lie group $G_3$ can be written in the form of exponent of a corresponding element of the Lie algebra $L_3$
$$
S = \exp(s) = e + s + \frac{s^2}{2!} + \cdots = e\cos|s| + \frac{s}{|s|}\sin|s|,
$$
where $s=s_k\tau^k\in L_3$ and $|s|=\sqrt{s_1{}^2+s_2{}^2+s_3{}^2}$.

Let us define a  Lie group $G(\chi)$ and its real Lie algebra $L(\chi)$
\begin{eqnarray*}
L(\chi) &=& \{s\in K(\chi)\,:\,s^\dagger=-s\},\\
G(\chi) &=& \{S\in \C\otimes\cl^\Even\,:\,  S-e\in K(\chi),\  S^\dagger S=e\}.
\end{eqnarray*}
It is easy to check that the Lie algebra $L(\chi)$ is a real four-dimensional vector space spanned by the basis elements $\tau^1\chi,\tau^2\chi,\tau^3\chi,\theta\chi$.
The Lie algebra $L(\chi)$ is isomorphic to the Lie algebra $\u(2)$ and the Lie group $G(\chi)$ is isomorphic to the Lie group $\U(2)$ of unitary matrices of second order.

If $s\in L_4$, then $r=s\chi\in L(\chi)$. Any element $S\in G(\chi)$ can be written in the form of exponent of a corresponding element $r\in L(\chi)$
\begin{align*}
S&=\exp(r)=\exp(s\chi)=e +(\exp(s)-e)\chi,\\
S^{-1}&=\exp(-r)=\exp(-s\chi)=e +(\exp(-s)-e)\chi.
\end{align*}

\section{Elements of Clifford Analysis in Minkowski Space}
\label{sec3}
\noindent{ \bf The Minkowski space ${\Bbb R}^{1,3}$ and tensor fields}. We use Cartesian coordinates $x^\mu$, $\mu=0,1,2,3$ in Minkowski space ${\Bbb R}^{1,3}$. The metric tensor of the Minkowski space is given by the diagonal matrix $\eta=\|\eta_{\mu\nu}\|=\|\eta^{\mu\nu}\|={\rm diag}(1,-1,-1,-1)$. $\partial_\mu=\partial/\partial x^\mu$ are partial derivatives.

Consider changes of coordinates from the pseudoorthogonal (Lorentz) group $\O(1,3)$
 \begin{equation}
 x^\mu\to\acute{x}^\mu=p^\mu_\nu x^\nu,\label{acute:x}
 \end{equation}
 where
 $$
 P=\|p^\mu_\nu\|\in \O(1,3)
 $$
 and
 $$
 \O(1,3)=\{P\in \mathrm{GL}(4,\R) : P^T\eta P=\eta\},
 $$
 $P^T$ is the transposed matrix.

We use real or complex tensor fields $u^{\mu_1\ldots\mu_k}_{\nu_1\ldots\nu_l}$ in $\R^{1,3}$. Under the change of coordinates (\ref{acute:x}) we have the transformation law
\begin{equation}
u^{\mu_1\ldots\mu_k}_{\nu_1\ldots\nu_l}\to
\acute{u}^{\mu_1\ldots\mu_k}_{\nu_1\ldots\nu_l}=
p^{\mu_1}_{\alpha_1}\cdots p^{\mu_k}_{\alpha_k}
q_{\nu_1}^{\beta_1}\cdots q_{\nu_l}^{\beta_l}
u^{\alpha_1\ldots\alpha_k}_{\beta_1\ldots\beta_l},\label{tens:rule}
\end{equation}
where
$$
Q=\|q_\nu^\beta\|=P^{-1}.
$$
The set of tensor fields with $k$ contravariant indices and $l$ covariant indices is denoted by
$\T^k_l$. We write $u\in\T^k_l$, or $u^{\mu_1\ldots\mu_k}_{\nu_1,\ldots,\nu_l}\in\T^k_l$.
\medskip

\noindent{\bf Tensor fields with values in Clifford algebras}. Also, we use tensor fields with values in Clifford algebras $\cl$, $\C\otimes\cl$ (components of tensor field $u^{\mu_1\ldots\mu_k}_{\nu_1\ldots\nu_l}$ are elements of $\cl$ or $\C\otimes\cl$). In this case we write  $u^{\mu_1\ldots\mu_k}_{\nu_1,\ldots,\nu_l}\in\cl\T^k_l$ (or in $\C\otimes\cl\T^k_l$).
If components of a tensor field $u$ belong to the subspace $\cl^r$ ($0\leq r\leq4$), then we write $u\in\cl^r\T^k_l$. If components of a tensor field $u$ belong to some Lie algebra $L\subset\C\otimes\cl$,  then we write $u\in L\T^k_l$.

With a tensor field $u^{\mu_1\ldots\mu_k}_{\nu_1,\ldots,\nu_l}\in\T^k_l$  one can associate a tensor field with values in $\cl^0$ by the rule  $u^{\mu_1\ldots\mu_k}_{\nu_1,\ldots,\nu_l}\to u^{\mu_1\ldots\mu_k}_{\nu_1,\ldots,\nu_l}e\in\cl^0\T^k_l$, where $e$ is the identity element of the Clifford algebra $\cl$.

\medskip

\noindent{\bf A tetrad field $y^\mu_a$ and a genvector field $h^\mu$.} A set of four real orthonormal vector fields $y^\mu_a$ ($a=0,1,2,3$) in $\R^{1,3}$ is called {\em a tetrad}. The orthonormality condition means
$$
y^\mu_a y^\nu_b\eta^{ab}=\eta^{\mu\nu}.
$$
With the aid of a tetrad $y^\mu_a$ we define a vector field
$$
h^\mu:=y^\mu_a e^a\in\cl^1\T^1_0
$$
such that
\begin{equation}
h^\mu h^\nu+ h^\nu h^\mu=2\eta^{\mu\nu}e,\quad \mu,\nu=0,1,2,3.\label{gen:cond}
\end{equation}

If a tetrad $y^\mu_a$ satisfies the condition
\begin{equation}
\partial_\mu y^\mu_0=0,\label{div:y0}
\end{equation}
then
the vector field $h^\mu=y^\mu_a e^a$ is called {\em a genvector field}\footnote{According to (\ref{gen:cond}), components of a genvector field $h^\mu$  (at any point $x\in\R^{1,3}$) can be considered as generators of the Clifford algebra $\cl$.}.

Condition (\ref{div:y0}) gives the following condition for the genvector $h^\mu$:
\begin{equation}
\partial_\mu(\pi^0(\beta h^\mu))=0.\label{pi0:betah}
\end{equation}
If $h^\mu$ is a genvector and $S\in G_3$, then $\acute h^\mu = S^{-1}h^\mu S$ is also a genvector.
\medskip

\noindent{\bf A differential operator $h^\mu\partial_\mu$.} In field equations we use a differential operator of first order
$$
\eth =h^\mu\partial_\mu,
$$
acting on scalar and tensor fields with values in $\cl$ or in $\C\otimes\cl$.
 Note that the square of this operator is equal to the d'Alembert operator multiplied by the identity element $e$
 $$
\eth^2=e\,\partial^\mu\partial_\mu=e\, \square.
$$
So, the operator $\eth$ can be considered as an analog of the Dirac operator.
\medskip

\noindent{\bf A system of Yang-Mills equations.} Let $G$ be a semisimple Lie group and $L$ be the real Lie algebra of the Lie group $G$. Consider a system of Yang-Mills equations\footnote{In this paper we consider several gauge fields and we denote potential and  strength of a Yang-Mills field by the same letter with different numbers of indices, for example, $a_\mu,a_{\mu\nu}$.} in $\R^{1,3}$
\begin{equation}
\partial_\mu a_\nu-\partial_\nu a_\mu - [a_\mu, a_\nu] = a_{\mu\nu},\quad
\partial_\mu a^{\mu\nu}- [a_\mu, a^{\mu\nu}] = j^\nu,\label{YM:aa}
\end{equation}
where $a_\mu\in L\T^0_1$, $a_{\mu\nu}\in L\T^0_2$, $j^\nu\in L\T^1_0$ and these tensor fields are depend on $x\in\R^{1,3}$.
The system of Yang-Mills equations is invariant under the gauge transformation (gauge symmetry)
\begin{equation}
a_\mu\to S^{-1}a_\mu S - S^{-1}\partial_\mu S,\quad a_{\mu\nu}\to S^{-1}a_{\mu\nu}S,\quad
j^\nu \to S^{-1}j^\nu S,\label{gauge:sym:0}
\end{equation}
where $S=S(x)\in G$.
Also, it is well known that Yang-Mills equations (\ref{YM:aa}) have the consequence
\begin{equation}
\partial_\nu j^\nu - [a_\nu, j^\nu] =0.\label{YM:conseq}
\end{equation}
The pair $(a_\mu,a_{\mu\nu})$ is called {\em a Yang-Mills field}; $a_\mu$ is the potential of Yang-Mills field and $a_{\mu\nu}$ is the strength of Yang-Mills field.

\section{The Main System of Equations with the $\SU(2)\times \U(2)$ Gauge Symmetry}
\label{sec4}
\sloppypar{
In this section we present a system of partial differential equations in Minkowski space with two unitary gauge symmetries. The first gauge group $\U(2)$ corresponds to the electroweak \footnote{In this paper we do not consider important details of electoweak theory concerning a left-handed property of interacted particles and the Higgs mechanism.} interaction of the
 Standard Model and the second gauge group $\SU(2)$ corresponds to gravitational interactions of leptons. In the next section we consider equations for quarks.
}
\medskip

\noindent{\bf The main system of equations.}
Let us consider a system of equations
\begin{eqnarray}
&& h^\mu(\partial_\mu\Psi+\Psi A_\mu-C_\mu\Psi)+i m\Psi=0,\label{main1bxy}\\
&& \partial_\mu A_\nu-\partial_\nu A_\mu-[A_\mu,A_\nu]=A_{\mu\nu},\label{main2bxy}\\
&& \partial_\mu A^{\mu\nu}-[A_\mu,A^{\mu\nu}]= \Psi^\dagger i\beta h^\nu\Psi,\label{main3bxy}\\
&& \partial_\mu C_\nu-\partial_\nu C_\mu-[C_\mu,C_\nu]=C_{\mu\nu},\label{main4bxy}\\
&& \partial_\mu C^{\mu\nu}-[C_\mu,C^{\mu\nu}]= \theta\beta h^\nu-\pi^4(\theta\beta h^\nu),\label{main5bxy}
\end{eqnarray}
 where $h^\mu$ is a genvector; a scalar field $\Psi=\Psi\chi$  belongs to the left ideal $I(\chi)\subset\C\otimes\cl$, which is generated by the Hermitian idempotent $\chi=(e- i\theta)/2$; $A_\mu\in L(\chi)\T_1^0$; $A_{\mu\nu}\in L(\chi)\T_2^0$; $C_\mu\in L_3\T_1^0$; $C_{\mu\nu}\in L_3\T_2^0$; $m$ is a real number (a mass of particle); $i$ is the imaginary unit. The quantities $h^\mu,\Psi,A_\mu,A_{\mu\nu},C_\mu,C_{\mu\nu}$  depend on $x\in\R^{1,3}$  and the quantities $\chi,\beta,\theta,m,i$ are independent of $x$.

The system of equations (\ref{main1bxy})-(\ref{main5bxy}) consists of the Dirac type equation (\ref{main1bxy}) \cite{paper2002}\footnote{the equation \cite{paper2002} can be considered as a development of the M.~Riesz's form of Dirac equation \cite{Riesz}.} for a wave function $\Psi$
and two Yang-Mills systems of equations --  (\ref{main2bxy}), (\ref{main3bxy}) and (\ref{main4bxy}), (\ref{main5bxy}) for Yang-Mills potentials  $A_\mu,C_\mu$ and for Yang-Mills strengths    $A_{\mu\nu},C_{\mu\nu}$.

We suppose that, after detailing the model, the Yang-Mills field $(A_\mu,A_{\mu\nu})$ can be interpreted as the gauge field of electroweak Standard Model  (with the $\U(1)\times \SU(2)$ gauge symmetry). And the Yang-Mills field $(C_\mu, C_{\mu\nu})$ can be  interpreted as a gauge field with $\SU(2)$ symmetry, which describes the gravitational interaction of elementary particles ($C_\mu$ is a potential of the gravitation field and  $C_{\mu\nu}$ is a strength of the gravitation field).

Let us show that the system of equation  (\ref{main1bxy})-(\ref{main5bxy}) is invariant under two gauge transformations (gauge symmetries) with Lie groups $G(\chi)$ and $G_3$.

\begin{theorem}
1) If $U=U(x)\in G(\chi)$, $x\in\R^{1,3}$ and tensor fields $\Psi$, $h^\mu$, $A_\mu,A_{\mu\nu}$, $C_\mu,C_{\mu\nu}$ satisfy equations (\ref{main1bxy})-(\ref{main5bxy}), then the tensor fields with the prime
\begin{align}
\acute\Psi &= \Psi U,\nonumber\\
\acute h^\mu &= h^\mu,\nonumber\\
\acute A_\mu &= U^{-1}A_\mu U-U^{-1}\partial_\mu U,\nonumber\\
\acute A_{\mu\nu} &= U^{-1}A_{\mu\nu}U,\label{rule:Gt}\\
\acute C_\mu &= C_\mu,\nonumber\\
\acute C_{\mu\nu} &= C_{\mu\nu}\nonumber
\end{align}
also satisfy equations (\ref{main1bxy})-(\ref{main5bxy}).

2) If $S=S(x)\in G_3$, $x\in\R^{1,3}$ and tensor fields $\Psi$, $h^\mu$, $A_\mu,A_{\mu\nu}$, $C_\mu,C_{\mu\nu}$ satisfy equations (\ref{main1bxy})-(\ref{main5bxy}), then the tensor fields with the prime
\begin{align}
\acute\Psi &= S^{-1}\Psi S,\nonumber\\
\acute h^\mu &= S^{-1}h^\mu S,\nonumber\\
\acute A_\mu &= S^{-1}A_\mu S-\chi S^{-1}\partial_\mu S,\nonumber\\
\acute A_{\mu\nu} &= S^{-1}A_{\mu\nu}S,\label{rule:G3}\\
\acute C_\mu &= S^{-1}C_\mu S -S^{-1}\partial_\mu S,\nonumber\\
\acute C_{\mu\nu} &= S^{-1}C_{\mu\nu}S\nonumber
\end{align}
also satisfy equations (\ref{main1bxy})-(\ref{main5bxy}).

 The quantities $\chi,\theta,\beta,i,m$ do not change under both transformations.
\end{theorem}

 The proof is by direct calculation.


\medskip

Let us denote the right hand parts of Yang-Mills equations (\ref{main2bxy})--(\ref{main5bxy}) by
$$
J^\mu_{(A)} := \Psi^\dagger i\beta h^\nu\Psi,\quad
J^\mu_{(C)} := \theta\beta h^\nu-\pi^4(\theta\beta h^\nu).
$$
For the Yang-Mills equations (\ref{main2bxy})--(\ref{main5bxy}) we have consequences (see (\ref{YM:conseq}))
\begin{eqnarray}
&& \partial_\mu J^\mu_{(A)} - [A_\mu, J^\mu_{(A)}]=0,\label{J:Ax1}\\
&& \partial_\mu J^\mu_{(C)} - [C_\mu, J^\mu_{(C)}]=0.\label{J:Cx1}
\end{eqnarray}

Let us calculate a consequence of the Dirac type equation (\ref{main1bxy}). Multiplying both sides of (\ref{main1bxy}) by $\Psi^\dagger i\beta$, we obtain
\begin{equation}
\Psi^\dagger i\beta h^\mu(\partial_\mu\Psi+\Psi A_\mu-C_\mu\Psi)-m\Psi^\dagger\beta \Psi=0.\label{PsiDir1b}
\end{equation}
Using formulas
$$
A_\mu^\dagger=-A_\mu,\quad C_\mu^\dagger=-C_\mu,\quad
(i\beta h^\mu)^\dagger=-i\beta h^\mu,\quad
(\theta\beta h^\mu)^\dagger=-\theta\beta h^\mu,
$$
we get the Hermitian conjugated equality
\begin{equation}
-(\partial_\mu\Psi^\dagger-A_\mu\Psi^\dagger+\Psi^\dagger C_\mu)i\beta h^\mu\Psi-m \Psi^\dagger\beta\Psi=0.\label{PsiDir2b}
\end{equation}
Subtracting (\ref{PsiDir2b}) from (\ref{PsiDir1b}), we see that the result can be written in the form
\begin{equation}
((\partial_\mu(\Psi^\dagger i\beta h^\mu\Psi)- [A_\mu,\Psi^\dagger i\beta h^\mu\Psi])
+\Psi^\dagger i\theta(\partial_\mu(\theta\beta h^\mu)-[C_\mu,\theta\beta h^\mu])\Psi=0.\label{1st:conb}
\end{equation}

 We claim that consequences (\ref{J:Ax1}), (\ref{J:Cx1}) of Yang-Mills equations  (\ref{main2bxy}), (\ref{main3bxy})  and (\ref{main4bxy}), (\ref{main5bxy}) are compatible with the consequence (\ref{1st:conb}) of the Dirac type equation (\ref{main1bxy}).

\begin{theorem}
If equalities (\ref{J:Ax1}), (\ref{J:Cx1}) are satisfied, then the equality (\ref{1st:conb}) is also satisfied.
\end{theorem}
Proof. Let us show that from the equality (\ref{J:Cx1}) we get the equality
\begin{equation}
\partial_\mu(\theta\beta h^\mu) - [C_\mu, \theta\beta h^\mu]=0.\label{thetabetah}
\end{equation}
In fact, we have
$$
\theta\beta h^\mu\in\cl^2\oplus\cl^4
$$
that means
$$
\theta\beta h^\mu=\pi^2(\theta\beta h^\mu) + \pi^4(\theta\beta h^\mu).
$$
Since $C_\mu\in\cl^2$, it follows that
\begin{equation}
[C_\mu, \pi^4(\theta\beta h^\mu)]=0.\label{Cmupi}
\end{equation}
According to (\ref{pi0:betah}) we obtain
\begin{equation}
\partial_\mu \pi^4(\theta\beta h^\mu)=\theta\,\partial_\mu \pi^0(\beta h^\mu)=0.\label{pi4J}
\end{equation}
So, from (\ref{Cmupi}), (\ref{pi4J}), (\ref{J:Cx1}) we obtain (\ref{thetabetah}).

From the formulas (\ref{J:Ax1}) and (\ref{thetabetah}) we see that the left hand part of (\ref{1st:conb}) is equal to zero. This completes the proof.


\section{The Main System of Equations with the $\SU(2)\times \U(2)\times \U(3)$ Gauge Symmetry}\label{sec5}
In this section we write down a system of equations for quarks, which is invariant under  three gauge transformations with Lie groups $\SU(2)$, $\U(2)$, and $\U(3)$.
Two gauge groups $\U(2)$, $\U(3)$ corresponds to the electroweak and to the strong (QCD) interactions of the Standard Model. The $\SU(2)$ gauge group corresponds to the gravitational interaction of quarks.

\medskip

Consider several mathematical structures.

1) Tensor and scalar fields in Minkowski space $\R^{1,3}$ with values in the complexified Clifford algebra $\C\otimes\cl$:  $h^\mu\in\cl^1\T^1_0$ is a genvector;
$\Psi_1,\Psi_2,\Psi_3\in I(\chi)$; $A_\mu\in L(\chi)\T^0_1$, $A_{\mu\nu}\in L(\chi)\T^0_2$; $C_\mu\in L_3\T^0_1$, $C_{\mu\nu}\in L_3\T^0_2$;

2) By the underlined letters we denote $3\!\times\!3$-matrices from $\Mat(3,\C)$: $\underline{T},\underline{E},\underline{B}_\mu\in\Mat(3,\C)$, where
$$
\underline{T}=\begin{pmatrix}1&0&0\cr 0&0&0\cr 0&0&0\end{pmatrix},
\quad
\underline{E}=\begin{pmatrix}1&0&0\cr 0&1&0\cr 0&0&1\end{pmatrix},
\quad
\underline{B}_\mu=\begin{pmatrix}b^1_{1\mu}&b^1_{2\mu}&b^1_{3\mu}\cr
b^2_{1\mu}&b^2_{2\mu}&b^2_{3\mu}\cr b^3_{1\mu}&b^3_{2\mu}&b^3_{3\mu}
\end{pmatrix},
$$

3) By letters with $\,\check{}\, $ we denote elements of the tensor product $\Mat(3,\C)\otimes\cl$
$$
\check \Psi=(\underline{T}\otimes e)\check\Psi=\begin{pmatrix}\Psi_1&\Psi_2&\Psi_3\cr 0&0&0\cr 0&0&0\end{pmatrix},
$$
$$
\check A_\mu=\underline{E}\otimes A_\mu=\begin{pmatrix}A_\mu&0&0\cr 0&A_\mu&0\cr 0&0&A_\mu\end{pmatrix},
$$
$$
\check B_\mu=\underline{B}_\mu\otimes e=\begin{pmatrix}b^1_{1\mu}e&b^1_{2\mu}e&b^1_{3\mu}e\cr
b^2_{1\mu}e&b^2_{2\mu}e&b^2_{3\mu}e\cr b^3_{1\mu}e&b^3_{2\mu}e&b^3_{3\mu}e
\end{pmatrix},
$$
$$
\check h^\mu=\underline{E}\otimes h^\mu,\quad
\check C_\mu=\underline{E}\otimes C_\mu,\quad
\check \beta=\underline{E}\otimes \beta,\quad
\check\chi=\underline{E}\otimes\chi.
$$

Now we may write down a Dirac type equation
\begin{equation}
\check h^\mu(\partial_\mu\check \Psi + \check \Psi \check A_\mu + \check \Psi \check B_\mu - \check C_\mu\check \Psi)+im\check \Psi=0.\label{U223:eq}
\end{equation}
This equation can be rewritten without matrices in the form of three equations in Clifford algebra $\C\otimes\cl$
\begin{equation}
h^\mu(\partial_\mu\Psi_l +\Psi_l A_\mu +\Psi_k b^k_{l\mu} -  C_\mu \Psi_l)+im\Psi_l=0,\quad l=1,2,3.\label{U223:eq1}
\end{equation}

We have three gauge symmetries of the equation (\ref{U223:eq}):

1) For $\check U=\underline{E}\otimes U$, $U\in G(\chi)$
\begin{align*}
\check\Psi &\to \check\Psi\check U,\\
\check h^\mu &\to \check h^\mu,\\
\check A_\mu &\to \check{U}^{-1}\check A_\mu\check U-  \check{U}^{-1}\partial_\mu\check{U}=
\underline{E}\otimes(U^{-1}A_\mu U-U^{-1}\partial_\mu U),\\
\check A_{\mu\nu} &\to \check{U}^{-1}\check A_{\mu\nu}\check U,\\
\check B_\mu &\to \check B_\mu,\\
\check B_{\mu\nu} &\to \check B_{\mu\nu},\\
\check C_\mu &\to \check C_\mu,\\
\check C_{\mu\nu} &\to \check C_{\mu\nu};
\end{align*}

2) For $\check V=\underline{V}\otimes e$, $\underline{V}\in \U(3)$
\begin{align*}
\check\Psi &\to \check\Psi\check V,\\
\check h^\mu &\to \check h^\mu,\\
\check A_\mu &\to \check A_\mu,\\
\check A_{\mu\nu} &\to \check A_{\mu\nu},\\
\check B_\mu &\to \check{V}^{-1}\check B_\mu\check V-  \check{V}^{-1}\partial_\mu\check{V}=
(\underline{V}^{-1}\underline{B}_\mu \underline{V}-\underline{V}^{-1}\partial_\mu \underline{V})\otimes e,\\
\check B_{\mu\nu} &\to \check{V}^{-1}\check B_{\mu\nu}\check V,\\
\check C_\mu &\to \check C_\mu,\\
\check C_{\mu\nu} &\to \check C_{\mu\nu};
\end{align*}

3) For $\check{S}=\underline{E}\otimes S$, $S\in G_3$
\begin{align*}
\check\Psi &\to \check{S}^{-1}\check\Psi\check S,\\
\check h^\mu &\to \check{S}^{-1}\check h^\mu\check{S}=\underline{E}\otimes(S^{-1}h^\mu S),\\
\check A_\mu &\to \check{S}^{-1}\check A_\mu\check S- \check{\chi}\, \check{S}^{-1}\partial_\mu\check{S}=
\underline{E}\otimes(S^{-1}A_\mu S- \chi\, S^{-1}\partial_\mu S),\\
\check A_{\mu\nu} &\to \check{S}^{-1}\check A_{\mu\nu}\check S,\\
\check B_\mu &\to \check B_\mu,\quad\mbox{($\check S$ commutes with $\check{B}_\mu$)},\\
\check B_{\mu\nu} &\to \check B_{\mu\nu},\\
\check C_\mu &\to \check{S}^{-1}\check C_\mu\check S- \check{S}^{-1}\partial_\mu\check{S}=
\underline{E}\otimes(S^{-1}C_\mu S- S^{-1}\partial_\mu S),\\
\check C_{\mu\nu} &\to \check{S}^{-1}\check C_{\mu\nu}\check S,
\end{align*}
 where $\check\chi =\underline{E}\otimes\chi$, $[\check\chi,\check S]=0$.
\medskip

Let $\check J^\mu$ be a vector field with values in the tensor product $\Mat(3,\C)\otimes\cl$
$$
\check J^\mu := \check \Psi^\dagger i\check \beta \check h^\mu\check \Psi = \|J^\mu_{kl}\|,
$$
where
$$
J^\mu_{kl} = \Psi^\dagger_k i\beta h^\mu\Psi_l.
$$
And let $\check\Psi^\dagger$ be the Hermitian conjugated element
$$
\check \Psi^\dagger=\check\Psi^\dagger(\underline{T}\otimes e)=\begin{pmatrix}\Psi_1^\dagger&0&0\cr\Psi_2^\dagger&0&0\cr \Psi_3^\dagger&0&0\end{pmatrix}.
$$
If we multiply the left hand side of  equation (\ref{U223:eq}) by
$\check\Psi^\dagger i\check\beta$ and subtract the Hermitian conjugated expression, then we obtain
\begin{equation}
(\partial_\mu \check J^\mu - [\check A_\mu+\check B_\mu,\check J^\mu])+\check \Psi^\dagger i\check \theta(\partial_\mu(\check \theta\check \beta \check h^\mu) - [\check C_\mu,\check \theta\check \beta \check h^\mu])\check \Psi=0.\label{1st:cons}
\end{equation}
Let us take the matrix trace operation $\tr\,:\,\Mat(3,\C)\otimes\cl\to\C\otimes\cl$ and define
$$
J^\mu_{(A)} :=\frac{1}{3}\tr\,\check J^\mu = \frac{1}{3}(J^\mu_{11}+J^\mu_{22}+J^\mu_{33}).
$$
Note that $\tr([\check B_\mu,\check J^\mu])=0$.
Applying the operation $\frac{1}{3}\tr$ to both sides of the equality $(\ref{1st:cons})$, we get the consequence
\begin{equation}
(\partial_\mu J_{(A)}^\mu - [A_\mu,J_{(A)}^\mu]) + \frac{1}{3}\tr\big(\check \Psi^\dagger i \check \theta(\partial_\mu( \check \theta\check \beta \check h^\mu) - [\check C_\mu,\check \theta\check \beta \check h^\mu])\check \Psi\big)=0.\label{2nd:cons}
\end{equation}
Consider the projection operator
$$
\pi^0\,:\,\Mat(3,\C)\otimes\cl\to\C\otimes\cl^0
$$
and an operator $\dot\pi^0\,:\,\Mat(3,\C)\otimes\cl\to\C$,
$$ \dot\pi^0(U)=\pi^0(U)|_{e\to1}.
$$
Then
$$
\underline{J}^\mu_{(B)}:=\dot\pi^0(\check J^\mu) = \begin{pmatrix}
\dot\pi^0(J^\mu_{11}) & \dot\pi^0(J^\mu_{12}) & \dot\pi^0(J^\mu_{13})\cr
\dot\pi^0(J^\mu_{21}) & \dot\pi^0(J^\mu_{22}) & \dot\pi^0(J^\mu_{23})\cr
\dot\pi^0(J^\mu_{31}) & \dot\pi^0(J^\mu_{32}) & \dot\pi^0(J^\mu_{33})
\end{pmatrix}\in\Mat(3,\C).
$$
It is not hard to prove that
$$
\dot\pi^0([\check B_\mu,\check J^\mu]) = [\underline B_\mu,\dot\pi^0(\check J^\mu)].
$$
Hence, applying the operator $\dot\pi^0$ to both parts of the equality (\ref{1st:cons}), we arrive at the equality
\begin{equation}
(\partial_\mu\underline{J}_{(B)}^\mu - [B_\mu,\underline{J}_{(B)}^\mu]) + \pi^0\big(\check \Psi^\dagger i \check \theta(\partial_\mu( \check \theta\check \beta \check h^\mu) - [\check C_\mu,\check \theta\check \beta \check h^\mu])\check \Psi\big)=0.\label{3d:cons}
\end{equation}

So, the following theorem is proved:
\begin{theorem}
If
\begin{eqnarray}
&& \partial_\mu J^\mu_{(A)} - [A_\mu, J^\mu_{(A)}]=0,\label{J:A}\\
&& \partial_\mu\underline{J}^\mu_{(B)} - [\underline{B}_\mu, \underline{J}^\mu_{(B)}]=0,\label{J:B}\\
&& \partial_\mu J^\mu_{(C)} - [C_\mu, J^\mu_{(C)}]=0,\label{J:C}
\end{eqnarray}
where
$$
J^\mu_{(A)} = \frac{1}{3}(J^\mu_{11} +J^\mu_{22} +J^\mu_{33}),\quad
\underline{J}^\mu_{(B)}=\dot\pi^0(\check J^\mu),\quad
J^\mu_{(C)} = \theta\beta h^\mu-\pi^4(\theta\beta h^\mu),
$$
and
$$
\check J^\mu= \begin{pmatrix}
J^\mu_{11} & J^\mu_{12} & J^\mu_{13}\cr
J^\mu_{21} & J^\mu_{22} & J^\mu_{23}\cr
J^\mu_{31} & J^\mu_{32}&  J^\mu_{33}
\end{pmatrix}=\check\Psi^\dagger i\check\beta\check h^\mu\check\Psi,
$$
then the equalities (\ref{2nd:cons}) and (\ref{3d:cons}) are satisfied.
\end{theorem}
The proof is straightforward (see Theorems 1 and 2).

Let us postulate that covectors $A_\mu,\underline{B}_\mu,C_\mu$ satisfy Yang-Mills equations with non-Abelian currents  $J^\nu_{(A)}, \underline{J}^\nu_{(B)}, J^\nu_{(C)}$
\begin{eqnarray}
\partial_\mu A_\nu - \partial_\nu A_\mu - [A_\mu, A_\nu] = A_{\mu\nu},\nonumber\\
\partial_\mu A^{\mu\nu} - [A_\mu, A^{\mu\nu}] = J^\nu_{(A)},\label{YM:A}\\
\partial_\mu\underline{B}_\nu - \partial_\nu\underline{B}_\mu - [\underline{B}_\mu, \underline{B}_\nu] = \underline{B}_{\mu\nu},\nonumber\\
\partial_\mu \underline{B}^{\mu\nu} - [\underline{B}_\mu, \underline{B}^{\mu\nu}] = \underline{J}^\nu_{(B)},\label{YM:B}\\
\partial_\mu C_\nu - \partial_\nu C_\mu - [C_\mu, C_\nu] = C_{\mu\nu},\nonumber\\
\partial_\mu C^{\mu\nu} - [C_\mu, C^{\mu\nu}] = J^\nu_{(C)},\label{YM:C}
\end{eqnarray}

From these equations we have consequences (\ref{J:A}), (\ref{J:B}), (\ref{J:C}). Therefore the Yang-Mills equations (\ref{YM:A}), (\ref{YM:B}), (\ref{YM:C})  are compatible with the consequence (\ref{2nd:cons}) of the Dirac type equation (\ref{U223:eq}).

So, we arrive at a system of field equations for quarks with $\SU(2)\times \U(2)\times \U(3)$ gauge symmetry, which consists of the Dirac type equation (\ref{U223:eq}) and three pairs of Yang-Mills equations (\ref{YM:A}), (\ref{YM:B}), (\ref{YM:C}).


\end{document}